\documentclass[seceq]{ptptex}

\usepackage{graphicx}

\notypesetlogo

\title{%
$\theta$-Vacuum
}
\subtitle{ Phase Transitions and/or Symmetry Breaking  
at $\theta = \pi$%
}

\author{%
Vicente \textsc{Azcoiti},$^1$ Angelo \textsc{Galante}$^2$ and
Victor \textsc{Laliena}$^1$
}

\inst{
 $^1$Departamento de F\'\i sica Te\'orica, Facultad 
de Ciencias, Universidad de Zaragoza, 50009 Zaragoza, Spain \\
 $^2$Dipartimento di Fisica dell'Universit\`a di L'Aquila, 67100 L'Aquila, Italy \\
and\\ 
INFN, Laboratori Nazionali del Gran Sasso, 67010 Assergi, L'Aquila, Italy
}

\abst{%
Assuming that a quantum field theory 
with a $\theta$-vacuum term in the action shows non-trivial 
$\theta$-dependence and provided that some reasonable properties of 
the probability distribution function of the order parameter hold,
we argue that the theory either breaks 
spontaneously $CP$ at $\theta = \pi$ or shows a singular behavior at some 
critical $\theta_c$ between 0 and $\pi$. This result, which applies to any 
model with a pure imaginary contribution to the euclidean action consisting 
in a quantized charge coupled to a phase, as QCD, is illustrated with two 
simple examples; one of them intimately related to Witten's result on $SU(N)$ 
in the large $N$ limit.
}

\begin{document}

\maketitle

\section{Introduction}
Quantum Field Theories with a topological term in the action are a
subject of interest in high energy particle physics and in solid state
physics since a long time. In particle physics, these models 
describe particle interactions with a $CP$ violating term. The extremely
small experimental bound for the $CP$ violating effects in QCD (strong $CP$ 
problem) is still waiting for a convincing theoretical 
explanation. \cite{strongcp} \  
In solid state physics, chains of half-integer quantum spins with
antiferromagnetic interactions are related to the two-dimensional $O(3)$
nonlinear sigma model with topological term at $\theta=\pi$. It has been 
argued that this model presents a second order phase transition at 
$\theta=\pi$, keeping its ground state $CP$ symmetric 
(Haldane conjecture). \cite{haldane} \ 

Non-perturbative studies of field theories 
with a $\theta$-vacuum term are enormously 
delayed because of two technical reasons: first, 
there are difficulties in finding a consistent 
definition of topological charge for QCD in discrete space-time, 
preventing 
the use of the most powerful non-perturbative method, the lattice 
regularization; second, even in simpler 
QCD-like models, as $CP^N$ or sigma models, for which a consistent definition 
of topological 
charge is known, the complex nature of the euclidean action forbids 
the application of all standard Monte-Carlo algorithms to perform numerical
simulations (see for example Refs.~\citen{plefka} and \citen{imachi}). 
We need therefore new ideas which, beside our present technical 
abilities, allow a progress in this field.

The $\theta$-term breaks explicitly a $Z_2$ symmetry ($CP$), except at
$\theta=0$ and $\theta=\pi$. Most of the models the $\theta$ dependence 
of which is known break this symmetry spontaneously at $\theta=\pi$
(for an early study see Ref.~\citen{Hetrick}). 
An intriguing question in such a case
is the realization of the $2\pi$ periodicity in $\theta$. \cite{periodicity} \  
For instance, in
$SU(N)$ gauge theory at large $N$ the $\theta$ dependence appears through the
combination $\theta/N$. The periodicity is realized by conjecturing that the 
vacuum energy density is a multi-branched function of the form

\begin{equation}
E(\theta)\;=\;N^2\,\min_k\,H\left (\frac{\theta+2\pi k}{N}\right )\, .
\label{vacuum}
\end{equation}

\noindent
This behavior occurs also in some two dimensional systems, 
\cite{coleman,CPN,monos} \  and leads to
spontaneous breaking of $CP$ at $\theta=\pi$ in a natural way. Recently,
Witten has 
shown that in $SU(N)$ gauge theories at large $N$ Eq.~(\ref{vacuum}) holds with
$H(\theta)=C \theta^2$, where $C$ is a positive constant independent of $N$.
\cite{WITTEN2} \ 

We want to show in this paper that the above behavior is not specific of large
$N$ gauge theories but rather general. More precisely, we will argue that 
{\it any
model with a quantized charge which is an order parameter of a given symmetry
and which appears as an imaginary contribution to the euclidean action, either
breaks the symmetry at $\theta=\pi$ or shows some singularity for 
$\theta$-values between 0 and $\pi$, provided that there is a nontrivial
$\theta$-dependence}. 

\section{ General description}
In the specific case of QCD, the partition function 
with a topological term in the action reads as follows:

\begin{equation}
{\cal Z}\;=\;\int\,[dA_\mu^a]\,[d\psi]\,[d\bar\psi]\,\exp\,
\left\{\,-\int\,d^4x\,
[{\cal L}(x)-i\theta X(x)]\,\right\},
\label{QCDpf}
\end{equation}

\noindent
where ${\cal L}(x)$ is the standard QCD Lagrangian and
$X(x)\;=\;\frac{1}{16\pi^2}\,\epsilon_{\mu\nu\rho\sigma}\,\mathrm {Tr}\,
F_{\mu\nu}F_{\rho\sigma}(x)$
is the euclidean local density of topological charge, the normalization 
of which has been chosen in such a way that the topological charge,
$\int\,d^4x\, X(x)$ be an integer. We will assume in what follows that the 
regularized theory preserves the quantization of the topological charge.

To start with the proof, let us write the partition function $Z_V (\theta)$ 
in a finite space-time 
volume, $V$, as a sum over all topological sectors, labeled by the integer
$n$ that gives the topological charge, of the partition functions of
each sector, weighted with the proper topological phase:

\begin{equation}
{\cal Z}_V(\theta)\;=\;\sum_n\,g_V(n)\, e^{i\theta n}\, ,
\end{equation}

\noindent
where 

\begin{equation}
g_V(n)\;=\;\int_{n}\,[dA_\mu^a]\,[d\psi]\,[d\bar\psi]\,\exp\left[-\int\,d^4x\,
{\cal L}(x)\right]
\end{equation}

\noindent
is the standard partition function computed over the gauge sector with
topological charge equal to $n$. The function $g_V(n)$ is, up to the 
normalization factor $\sum_n g_V(n)$, the probability $p_V(n)$ of
the topological sector $n$ at $\theta=0$.
If we define the mean topological charge density as $x_n=n/V$,
we can write the previous partition function as

\begin{equation}
{\cal Z}_V(\theta)\;=\;\sum_{x_n}\,h_V(x_n)\,
e^{i\theta V x_n} \, ,
\label{pfdiscrete}
\end{equation}

\noindent
where $h_V(x_n)=g_V(n)$ and the step $\Delta x_n$ in the
topological charge density is $1/V$.

We will try now to write the partition function $Z_V (\theta)$ as a sum of 
{\it ``continuous''} pseudo-partition functions, 
where the word continuous stands to 
indicate that the density of topological charge $x$ entering the definition
of this new pseudo-partition function will 
be a continuous variable. To this end, 
let the new $h_V(x)$ be any continuous smooth interpolation of
$h_V(x_n)$ and let us define a new function of $\theta$ in the
following way:

\begin{equation}
{\cal Z}_{c,V}(\theta)\;=\;\int\,dx\,h_V(x)\,
e^{i\theta Vx} \, .
\label{pfcontinuum}
\end{equation}

Summing up the pseudo-partition functions 
${\cal Z}_{c,V}(\theta+2\pi m)$ for all integers $m$ 
and using the following representation of the periodic
delta function (Poisson summation formula):

\begin{equation}
\sum_m\,e^{i2\pi mVx}\;=\;\frac{1}{V}\,
\sum_m\,\delta\left(x-\frac{m}{V}\right) \, ,
\end{equation}

\noindent
we get the following identity: 

\begin{equation}
{\cal Z}_V(\theta)\;=\;V\,
\sum_m\,{\cal Z}_{c,V}(\theta+2\pi m) \, ,
\label{vacios}
\end{equation}

\noindent
which relates our QCD partition function ${\cal Z}_V(\theta)$
to the pseudo-partition functions ${\cal Z}_{c,V}(\theta+2\pi m)$.
Of course different interpolations of $h_V(x_n)$ will give different 
equivalent realizations of Eq.~(\ref{vacios}).

In deriving the above result we have assumed that the sum in $m$
and the integral in $x$ can be commuted. This is a working hypothesis 
on the properties of the $h_V(x_n)$ function and
we will use it to derive some physical consequences.
As will be evident in the next section, the validity of this assumption 
is confirmed in some examples where analytical calculations are possible.
Notice also that the sum in (\ref{vacios}) has nothing to do with the sum 
over different topological sectors. Each sector labeled by an integer 
$m$ in (\ref{vacios}) can get contributions from all topological sectors.
If ${\cal Z}_{c,V}(\theta+2\pi m)$ is positive for any integer $m$, the
physical interpretation of these integer numbers is very simple. Indeed 
in such a case and in what concerns the infinite volume limit, 
the summation in (\ref{vacios}) can be replaced by the 
maximum in $m$ of ${\cal Z}_{c,V}(\theta+2\pi m)$ (saddle point solution) and 
therefore the integers $m$  
will define metastable vacua which will decay to the true vacuum, the energy 
density of which is 
given by $\min_m\{-1/V \log {\cal Z}_{c,V}(\theta + 2\pi m)\}$ (see the first 
example discussed below).

The mean value of the density of topological charge can be written as

\begin{equation}
\langle x\rangle\;=\;\sum_m\,\langle x\rangle_{c,m}\,
\frac{{\cal Z}_{c,V}(\theta+2\pi m)}
{\sum_n{\cal Z}_{c,V}(\theta+2\pi n)} \, ,
\label{avtopcharge}
\end{equation}

\noindent
with

\begin{equation}
\langle x\rangle_{c,m}\;=\;
\frac{\int\,dx\,x\,h_V(x)\,
e^{i(\theta+2\pi m)Vx}}
{\int\,dx\,h_V(x)\,
e^{i(\theta+2\pi m)Vx}} \, .
\label{contitopch}
\end{equation}

\noindent
Since $x$ in Eqs. (\ref{pfcontinuum}) and (\ref{contitopch}) is a continuous 
variable, 
the pseudo-partition function $Z_{c,V}(\theta)$ will not be  a periodic 
function of $\theta$ and $\langle x\rangle_{c,m}$ needs not to vanish at 
$\theta = \pi$. The periodicity in $\theta$ of ${\cal Z}_{V}(\theta)$ is 
however guaranteed because we must sum up $Z_{c,V}(\theta+2\pi m)$ 
for all integer values of $m$. 
In the same way, $CP$ is recovered as a symmetry of the action at 
$\theta = \pi$, because the $m=0$ contribution to $\langle x\rangle$ in 
(\ref{avtopcharge}) 
is, at $\theta = \pi$, compensated by the $m=-1$ contribution, the $m=1$ 
contribution compensates with $m=-2$ and so on. Therefore, the contributions
of different sectors cancel each other, giving a vanishing density of 
topological charge at $\theta = \pi$. In the thermodynamic limit, 
however, one sector could dominate (see below) and this would imply that 
the limits 
$\theta\rightarrow\pi$ and $V\rightarrow\infty$ do not commute, giving rise
to the spontaneous $Z_2$ breaking.
We cannot exclude that $\langle x\rangle_{c,m}$ vanishes, by accident, 
for some integer 
$m$ at $\theta = \pi$. Usually however it is assumed that if the 
mean value of a local operator vanishes, it is because some symmetry forces  
it to vanish. In our case,
as previously discussed, the ``action'' in the pseudo-partition functions 
$Z_{c,V}(\theta+2\pi m)$ is not $CP$ invariant at $\theta = \pi$.

Equation (\ref{vacios}) is valid for any value of $\theta$. 
At $\theta=0$ (or at imaginary values of $\theta$)
and in the infinite volume limit the 
sector $m=0$ in (\ref{vacios}) gives all 
the contribution to the partition function.
This is a simple feature which follows from the fact that in this case 
all the terms in the summation (\ref{pfdiscrete}) and the integrand in 
(\ref{pfcontinuum}) are positive definite and then the partition functions 
(\ref{pfdiscrete}) and (\ref{pfcontinuum}) can be replaced, in the infinite 
volume limit, by the maximum of the summands and integrand respectively 
(saddle point solution). 
Since both maxima are coincident, we get the desired result.

Even if we cannot exclude on theoretical grounds a phase transition in
QCD at $\theta = 0$, it seems not very likely. Then one expects that
$m=0$ sector dominates at least in some interval around $\theta = 0$.
If this sector dominates for every $\theta$ between $-\pi$ and $\pi$,
we will get a phase transition at $\theta=\pi$, with eventually a
non-vanishing value of the topological charge density (remember the
discussion following Eq. (\ref{contitopch})) i.e. the theory will show
spontaneous $CP$ breaking. If by accident the density of topological
charge computed in the $m=0$ sector vanishes at $\theta = \pi$, we
should get a continuous phase transition since in any case sector $m =
-1$ will be the dominant one for $\theta$ values between $\pi$ and
$3\pi$.  The only way to avoid a phase transition at $\theta = \pi$
would be that the vacuum energy density of the $m=0$ sector were a
periodic function of $\theta$ with period equal to $2\pi$.  This of
course is what happens in some trivial cases as QCD in the chiral
limit, a limit in which there is no $\theta$-dependence, or the 
dilute instanton gas approximation. \cite{dilute} \ 
The last case is a model of 
non-interacting degrees of freedom, the partition function of which 
factorizes as the product of $V$ identical partition functions; and 
it is well known that phase transitions and critical phenomena are 
collective phenomena which appear as a consequence of the interaction of 
infinite degrees of freedom. Systems with non-interacting degrees of freedom 
are equivalent to systems with one degree of freedom and therefore they cannot show phase transitions. Therefore we can say that, excluding trivial 
cases, periodicity of the $m=0$ sector is not plausible.

The other possibility is that $m = 0$ sector dominates only until some 
critical $\theta_c$ less than $\pi$ and then 
other sectors start to give a contribution to the partition
function, in which case we will get a phase transition at this
$\theta_c$.

Since in the previous argumentation we have not made use of any specific
property of QCD, except the quantization of the topological charge, 
our result applies to any model with a quantized 
charge which appears as an imaginary contribution to the euclidean action.

\section{Two simple examples}
To see how this mechanism works in practical cases and to get
intuition of what we can expect in physical systems, let us analyze
in the following two simple examples: 
a model in which the probability distribution function of the density 
of topological charge at $\theta=0$ is assumed to be gaussian and
the Ising model within an imaginary external magnetic field.

The first example we want to discuss here 
includes models, as the quantum rotor, \cite{rotor} \  with a density of 
topological charge at $\theta=0$ distributed according to a gaussian 
distribution. Thus, let us assume that
the function $h_V(x_n)$ which enters Eq. (\ref{pfdiscrete}) 
has the form

\begin{equation}
h_V(x_n)\;=\;e^{-V a x_n^2} \, ,
\end{equation}

\noindent
where $a$ is a parameter related to the width of the 
distribution. This form of $h_V(x_n)$ is a natural assumption
from a physical point of view as a first approximation to the
actual distribution of nearly any model. In fact, outside
second order phase transitions, the probability distribution 
function of intensive operators as the density of topological
charge is expected to be gaussian in the vicinity of its maximum.
Of course, deviations from the gaussian behavior far from the
maximum can induce important changes in the $\theta$-dependence
of the theory in the large $\theta$ regime (as will 
become clear later). However, the gaussian
distribution provides us with a simple model that can be
analytically solved and gives useful insights on the general
problem that we are addressing.

The partition function of the model is 

\begin{equation}
{\cal Z}_V(\theta)\;=\;\sum_{x_n}\,e^{-Vax_n^2}
e^{i\theta Vx_n} \, .
\end{equation}

Even if this model can be solved without the help of Eq.~(\ref{vacios}), 
we will use it in order to clarify the relative weights of different 
sectors. 
The pseudo-partition function ${\cal Z}_{c,V}(\theta+2\pi m)$
entering Eq. (\ref{vacios}) can be analytically computed, and the final result 
is:

\begin{equation}
{\cal Z}_V(\theta)\;=\;\left(\frac{\pi V}{a}\right)^{1/2}\,
\sum_m\,e^{-\frac{1}{4a}(\theta+2\pi m)^2 V} \, .
\end{equation} 

\noindent
If $|\theta|<\pi$, it is simple to verify that the free 
energy density ${f}(\theta)$ is given by 
${f}(\theta)\;=\;\lim_{V\rightarrow\infty}\,
\frac{1}{V}\,\log\,{\cal Z}_V(\theta)\;=\;-\frac{1}{4a}\,
\theta^2 \,$. 
The $m=0$ sector dominates for every $\theta$ between $-\pi$ 
and $\pi$. The vacuum expectation value of the density of
topological charge, $\langle x\rangle$, is 

\begin{equation}
\langle x\rangle\;=\;i\frac{\theta}{2a} \qquad |\theta|<\pi
\, ,
\label{tcgaussian}
\end{equation}

\noindent
and the model breaks spontaneously parity at $\theta=\pi$ (see Fig. 1).

\begin{figure}[t!]
\centerline{\includegraphics*[width=3in,angle=90]{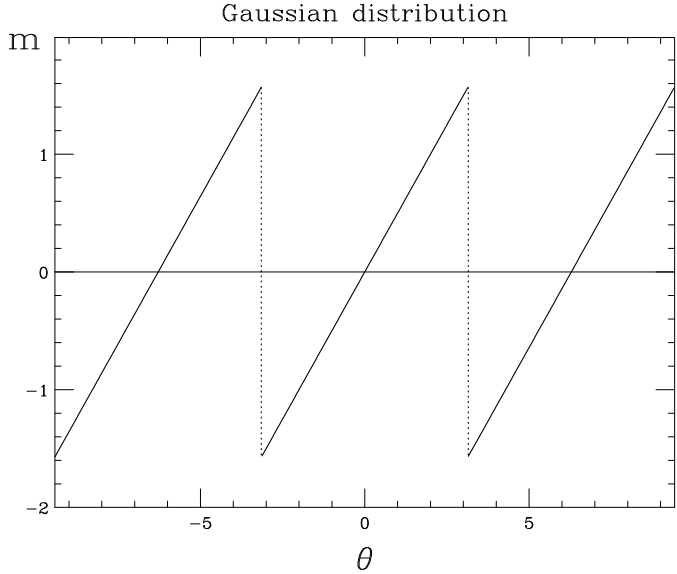}}
\caption{Expectation value of the topological charge density as a function of 
$\theta$ in models with gaussian distribution.}
\label{fig:gaussian}
\end{figure}  

This simple case that, as we have seen, illustrates very
well the results obtained in the first part of the paper, 
has also further relevance. 
In fact, using a duality
relation of gauge theories with a large number of colors
and string theory on a certain space-time manifold, Witten
has recently studied the $\theta$ dependence of pure gauge
theories in four dimensions \cite{WITTEN2} and found Eq.
(\ref{tcgaussian}). 
The combination of both results strongly suggests 
that the probability distribution function of the topological
charge density in pure $SU(N)$ gauge theory should be gaussian in
the large $N$ limit.

The second illustrative example is 
the one-dimensional Ising model 
within an imaginary external magnetic field. The hamiltonian of this model 
can be written as

\begin{equation}
H_N\;=\;-\,J\,\sum_{i=1}^N\,S_i\,S_{i+1}\:-\:i\,
\frac{\theta k_B T}{2}\,\sum_{i=1}^N\,S_i \, ,
\end{equation}

\noindent
where $J$ is the coupling constant between nearest neighbors,
$k_B$ is the Boltzmann constant, $T$ the physical temperature, $N$ the 
number of spins, and we assume periodic boundary conditions.
The partition function is given by

\begin{equation}
{\cal Z}_N\;=\;\sum_{\{S\}}\,e^{F\sum_i S_iS_{i+1}\,+\,
i\theta\frac{1}{2}\sum_i S_i},
\label{pfising}
\end{equation}

\noindent
where $F=J/k_B T$ and the sum is over all spin configurations.

For an even number of spins, the quantity $1/2\,\sum_i S_i$ which
appears in the imaginary part of the hamiltonian is an integer
taking values between $-N/2$ and $N/2$, and therefore it can
be seen as a quantized charge. Furthermore, the 
theory has a $Z_2$ symmetry at $\theta=0$ and $\theta=\pi$ which,
in the spirit of this work, is the analogue of $CP$ in QCD.

The transfer matrix technique allows to compute exactly the 
partition function defined in Eq.~(\ref{pfising}):
${\cal Z}_N=\lambda_+^N+\lambda_-^N$,
where $\lambda_\pm$ are the two eigenvalues of the transfer matrix

\begin{equation}
\lambda_\pm(\theta)\;=\;e^F\,\cos\frac{\theta}{2}\:\pm\:
\left(-e^{2F}\,\sin^2\frac{\theta}{2}\,+\,e^{-2F}
\right)^{1/2} \, 
\label{eigenvalues}
\end{equation}

\noindent
and we get for the mean value of the density of 
magnetization 

\begin{equation}
\langle x\rangle\;=\;i\,
\frac{\sin(\theta/2)}
{[e^{-4F}\,-\,\sin^2(\theta/2)]^{1/2}}, \qquad |\theta |<\pi \, .
\label{magnet}
\end{equation}

\noindent
The solution for every $\theta$ between $-\pi$ and $\pi$ is 
the analytical continuation of the solution for a real magnetic 
field.\cite{STANLEY} \  
Since in the real magnetic field case the $m = 0$ sector 
dominates always, 
this proves that the $m = 0$ sector dominates 
for every $\theta$ between $-\pi$ 
and $\pi$. 
A rather simple but long and straightforward calculation based on 
the analytical computation of the probability distribution function of the 
density of magnetization allows to demonstrate also this result, but the 
calculation is too long to be included here.

For $\theta$-values between $\pi$ and $3\pi$ 
($-\pi$ and $-3\pi$) the $m = -1$ ($m = 1$) sector dominates ($\lambda_-$ 
becomes larger than $\lambda_+$ in absolute value) 
and so on. Then 
we get, as expected, a periodic solution
which shows a first order phase transition at $\theta = \pi$ 
with spontaneous $Z_2$ breaking (see Fig. 2). 

\begin{figure}[t!]
\centerline{\includegraphics*[width=3in,angle=90]{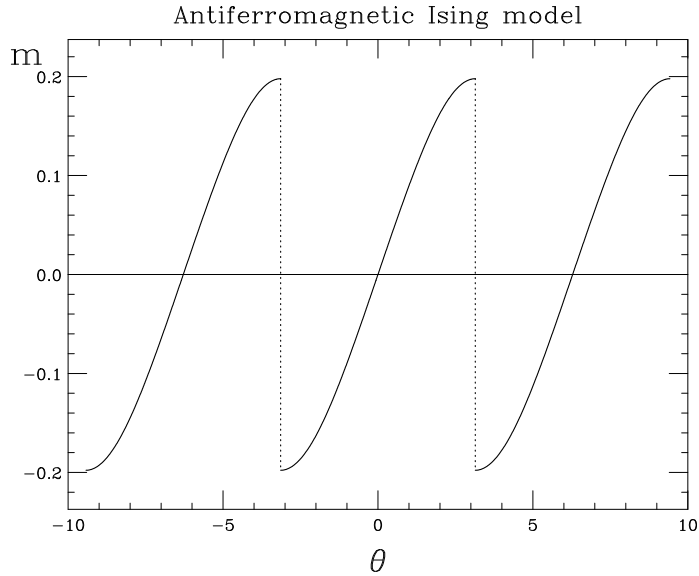}}
\caption{Magnetization as a function of 
$\theta$ in the one-dimensional antiferromagnetic Ising model.}
\label{fig:ising_af}
\end{figure}  

In the ferromagnetic case $(F\geq 0)$, the order parameter shows a 
divergence at a critical $\theta_{c}(T)$ running from 0 to 
$\pi$ \cite{PRL,CANTON}
and the model is ill defined for $\theta_{c}(T) < \theta < \pi$.

\section{Discussion}

Our main result, i.e., the existence of some singularity at $\theta_c$ 
between $0$ and $\pi$, was conjectured for non-abelian gauge theories by 
't Hooft \cite{HOOFT} in 1981. 
Based on the applicability of Poisson summation formula, it is
verified by the two simple examples previously 
discussed and also by other models with well established results as $CP^{N}$ 
models in the strong coupling region,\cite{CPN,monos} \  
two dimensional QED, \cite{coleman} \  or high temperature 
QCD at imaginary chemical potential,\cite{WEISS,PHILIP} \  
where the imaginary chemical potential 
plays the role of $\theta$ and the quantized charge is the baryonic charge.
However it still seems 
to be in contradiction with some predictions of 
chiral effective models of QCD.\cite{CHIRALEFF1, CHIRALEFF2} \ 
In these models the vacuum energy 
density is obtained, in the semi-classical approximation, 
as the solution of a given equation and the multiplicity 
of the solutions of this equation depends crucially on a certain inequality 
between quark masses. For realistic quark masses the solution would be unique, 
with periodicity $2\pi$, and the vacuum energy density would be an analytic 
function of $\theta$ for every $\theta$, 
including $\theta = \pi$.\cite{CHIRALEFF2} \    

There are however several possible explanations for the apparent 
contradiction between the chiral effective lagrangian results and ours, 
and we want just to enumerate some of them here. First the chiral effective 
lagrangian does not belong to the general class of models to which our results 
should apply since the quantized charge does not appear in it as an imaginary 
contribution to the euclidean action. Second the vacuum energy density 
obtained from the chiral effective lagrangian is the classical solution, i.e., 
it does not take into account quantum fluctuations which could change the 
vacuum structure; and third, higher order corrections to the chiral effective 
lagrangian could also change the vacuum energy density even at the classical 
level (see for instance Ref.~\citen{SMILGA}).

We want to stress also that the results reported 
in this article are complementary to the general analysis on spontaneous 
$CP$ breaking developed by Creutz in Ref.~\citen{CREUTZ}. \  In fact 
our results would exclude 
some general ansatzes for the effective potential compatible with the 
symmetries considered by Creutz in Ref.~\citen{CREUTZ} \  which drive to an order 
parameter analytical in $\theta$.

\section*{Acknowledgements}

V. Azcoiti thanks Philippe de Forcrand, Gerardo Ortiz and Owe Philipsen 
for comments and discussions. 
This work has been partially supported by an INFN-CICyT collaboration and 
MCYT (Spain), grant FPA2000-1252. 
Victor Laliena has been supported by Ministerio de Ciencia y Tecnolog\'{\i}a 
(Spain) under the Ram\'on y Cajal program.

\end{document}